\documentclass[aps,fleqn,twocolumn,prl,reprint]{revtex4-1}
\usepackage{graphicx}
\usepackage{amsmath}
\usepackage{amsbsy}

\usepackage[hidelinks]{hyperref}
\usepackage[charter]{mathdesign}

\def\sec#1{\emph{#1.}}

\def\d{{\rm d}}
\def\eps{\varepsilon}

\def\a{$\boldsymbol{\alpha}$}
\def\b{$\boldsymbol{\beta}$}
\def\c{$\boldsymbol{\gamma}$}

\def\Cavedef{\overline{C}}
\def\Cave{\overline{\mathbf{C}}}

\newif\ifall
\alltrue
\begin{document}
\title{Purifying electron spectra from noisy pulses with machine learning\\
 using synthetic Hamilton matrices}
\author{Sajal Kumar Giri}
\author{Ulf Saalmann}
\author{Jan M. Rost}
\affiliation{Max-Planck-Institut f{\"u}r Physik komplexer Systeme,
 N{\"o}thnitzer Str.\ 38, 01187 Dresden, Germany}
\date{\today}

\begin{abstract}\noindent
Photo-electron spectra obtained with intense pulses generated by free-electron lasers through self-amplified spontaneous emission are intrinsically noisy and vary from shot to shot.
We extract the purified spectrum, corresponding to a Fourier-limited pulse, with the help of a deep neural network. 
It is trained on a huge number of spectra, which was made possible by an extremely efficient propagation of the Schr\"odinger equation with synthetic Hamilton matrices and random realizations of fluctuating pulses. 
We show that the trained network is sufficiently generic such that it can purify atomic or molecular spectra, dominated
by resonant two- or three-photon ionization, non-linear processes which are particularly sensitive to pulse fluctuations. This is possible without training on those systems.
\end{abstract}

\pacs{32.80.Rm, %	Multiphoton ionization and excitation to highly excited states
 41.60.Cr}	% Free-electron lasers

\ifall\maketitle\fi

\ifall\relax\else\clearpage\fi

\noindent
Recent years have seen an avalanche-like increase of machine-learning applications in physics \cite{dubr18,mebu+19,caci+19}, which roughly fall into three categories: 
(a) applications within theory, e.g., for quantum information \cite{dubr18} or to elucidate intricate many-body properties \cite{catr17},
(b) within experiment to optimize experimental conditions, e.g., to characterize a free-electron laser (FEL) pulse \cite{sami+17}, 
and (c) applications that condition learning algorithms theoretically with the goal to apply the trained model to experimental data.
Our work falls in category (c). Although in principle far more general, we choose to be specific and apply the approach we develop to the purification of noisy photo-electron spectra as routinely obtained with self-amplified spontaneous emission (SASE) FELs operating in the desired frequency range.

\begin{figure}[t]\centering
\includegraphics[width=\columnwidth]{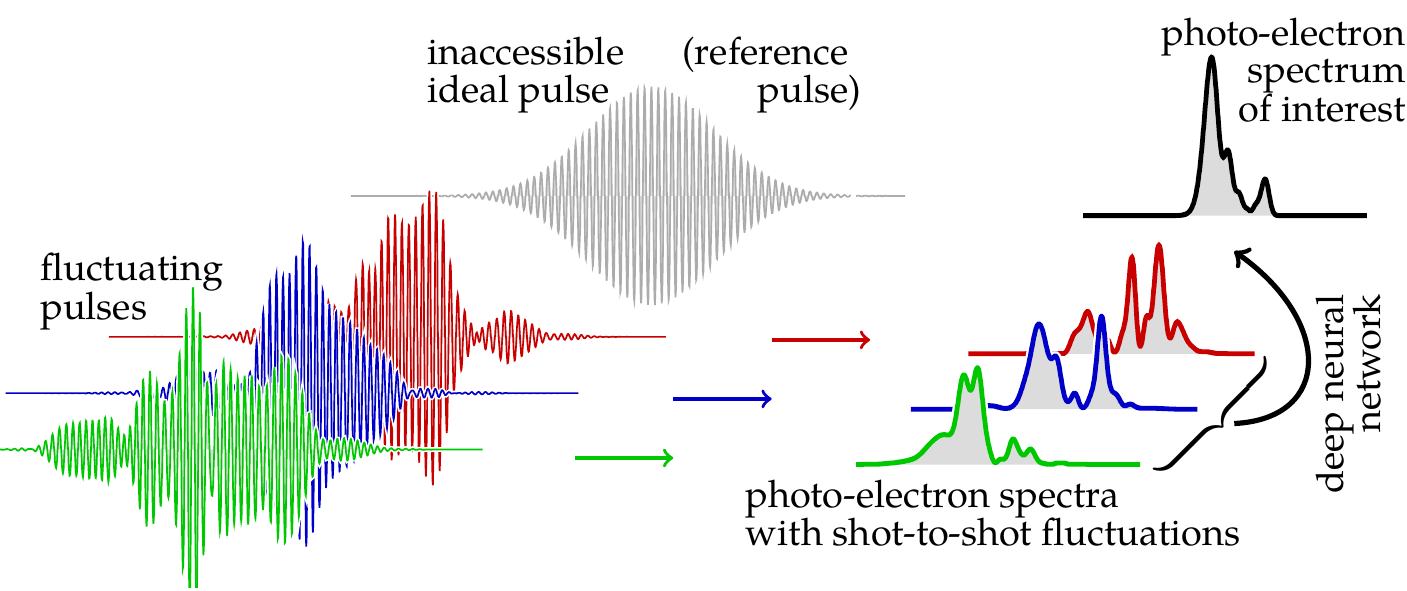}
\caption{Sketch of the problem: Photo-electron spectra
from fluctuating pulses are purified using a deep neural network.}
\label{fig:principle}
\end{figure}

Our goal is to train a deep neural network with sufficiently many noisy spectra and their pure counterparts, such that the trained network will be able to purify a noisy spectrum which is not contained in the training data, in particular an experimental one.
With purification, we mean that upon feeding with a noisy photo-electron spectrum the network returns a reference spectrum that would be obtained if the target system would be driven by an ideal Gaussian laser pulse, which we call the reference pulse, cf.\ Fig.\,\ref{fig:principle}.
This may seem straightforward. 
Yet, it is anything but trivial to generate a sufficient amount of suitable training data with an acceptable effort. 
This is in general the bottleneck for machine-learning applications in theory which requires new ways of thinking. 
In this vein, we introduce synthetic Hamilton matrices (SHMs). 
Synthetic means that we vary the matrix elements (here in a random fashion) about base values such that later on the trained network is able to purify real spectra from either experiment or theory.
The SHMs are constructed to speed up the generation of training data and we also expect them to become useful for other dynamical problems for which neural networks must be trained. Since the SHMs cover a large range of possible systems we can afford to use
for the base itself explicitly calculated photo-ionization dynamics in one dimension which is fast to compute and provides a suitable anchor point for the SHMs.

\sec{Setting up networks with SHMs}
To put our approach to a credible test we need
(i) a physical process, which is sensitive to the pulse profile,
(ii) a realistic way to model fluctuating pulses and we need to prepare a large set of spectra suitable for training the network. This involves (iii) a scheme to efficiently propagate millions of time-dependent Schr\"odinger equations,
(iv) a broad and uniform sampling of the generated spectra and
(v) a trainable parametrization. 
 %%%%%%%%%%%%%%%%%%%%%%%%%%%
%%%%%%%%%%%%%
 
(i) As a physical process which is non-linear in the driving light and therefore very sensitive to the intensity of the light pulse and hence its profile in time,
we have chosen quasi-resonant few-photon ionization. 
It can lead to multi-peak structures in the photo-electron spectrum \cite{auto55,role86,meen94,dece12a,basa+17}.

(ii) Fluctuating pulses from SASE FELs can be modeled by the so-called partial-coherence method \cite{pfji+10,mopf+11}, an experimentally verified method, which allows one to create ensembles of pulses $f_{l}(t)$ which differ through fluctuations 
while the ensemble average converges to a well defined pulse shape \cite{suppl}.
Those pulses have a characteristic duration $T$ and a coherence time $\tau$,
we use $T\,{=}\,3$\,fs and $\tau\,{=}\,1/2$\,fs here.
Apart from the intrinsic noise the pulses additionally jitter in their pulse energy.
We normalize all pulses $f_{l}(t)$ to unit pulse energy. This is also possible in the experiment as pulse energies can be easily measured shot-to-shot with gas monitor detectors \cite{tife+08}.

(iii) In principle numerical codes are available to propagate the time-dependent Schr\"odinger equation (TDSE) for one active electron in a strong laser field and calculate the resulting photo-electron spectrum $P(E)$ \cite{tasc12sc12,moba16,pamu16}.
However, the creation of a training data set from millions of pulses is prohibitively expensive, yet essential for successful deep-learning. 

To overcome this obstacle we work with Hamilton matrices whose construction is detailed in the supplement \cite{suppl}.
The new element, particularly formulated for the present context is 
the generation of $n_{\rm mat}$ Hamilton matrices with random energies $E_{\alpha}^{k}$, coupling matrix elements $V_{\alpha\beta}^{k}$, and field strengths $A_{k}$, corresponding to intensities (referring to the Fourier-limited pulse) in the range of $5{\times}10^{15}\!\ldots 5{\times}10^{16}$\,W/cm$^2$. Furthermore, for each Hamilton matrix the coupling to the light is augmented by $n_{\rm pul}$ noise realizations $f_{l}(t)$ with a central frequency of 21\,eV to arrive at
\begin{subequations}\label{eq:1d-hama}\begin{align}
\mathbf{E}_{k} & =E_{\alpha}^{k}\delta_{\alpha\beta},\quad\mathbf{V}_{k}=V_{\alpha\beta}^{k},
\\
\mathbf{H}_{kl}(t) & = \mathbf{E}_{k}+A_{k}\,f_{l}(t)\,\mathbf{V}_{k},
\end{align}\end{subequations}
whereby $k\,{=}\,1\ldots n_{\rm mat}$ and $l\,{=}\,1\ldots n_{\rm pul}$.
Boldface symbols in Eqs.\,\eqref{eq:1d-hama} describe matrices in terms of field-free states. 
It is only through these synthetic Hamilton matrices that we are able to create a sufficient number of non-trivial training data.
The matrices have been derived varying a 1D Hamilton operator (our base system), but since the energies $\mathbf{E}_{k}$ and the coupling matrix elements $\mathbf{V}_{k}$ are chosen randomly, these SHMs can purify real (3D) spectra, as we will see subsequently.

(iv) We have to create a set of spectra for training, validation and testing, which should cover to a large extent the domain of realizable spectra. This step is crucial and most expensive numerically, particularly when compared to the (modest) resources needed to set up and train the network.
%%%%%%%%%%%%%%%%%
\begin{figure}[t]
\centering
\includegraphics[width=\columnwidth]{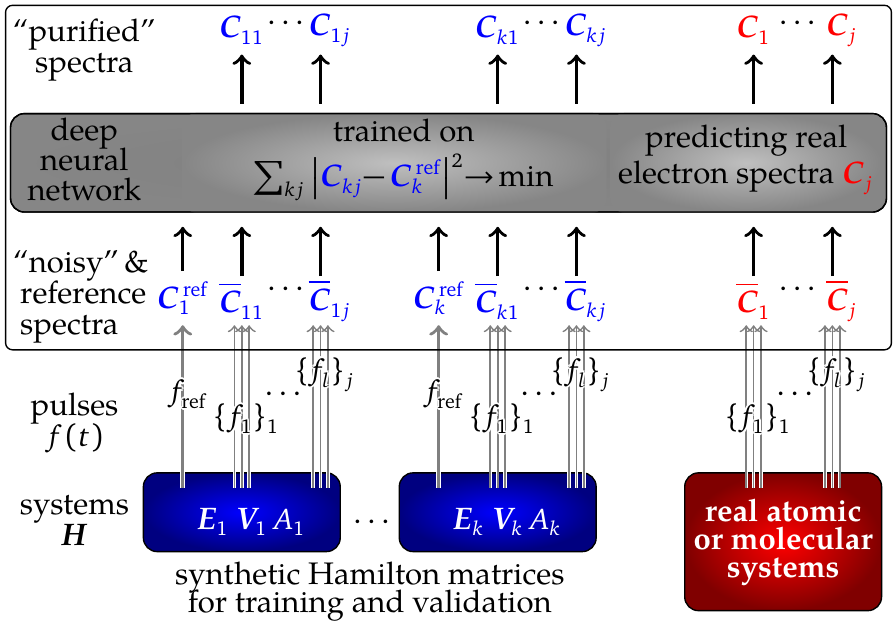}
\caption{Building a network with synthetic Hamilton matrices and noisy pulses (left-hand side, blue) and using it (right-hand side, red). The setup of $n_\mathrm{mat}$ SHMs is exemplified with the
first ($\mathbf{H}_{1}$) and $k$th ($\mathbf{H}_{k}$) one. With noisy pulse shapes $f_{l}(t)$ or the reference pulse shape $f_\mathrm{ref}(t)$ the SHMs are amended to $\mathbf{H}_{1l}$ and $\mathbf{H}_{kl}$, respectively, see Eqs.\,\eqref{eq:1d-hama}. Noisy spectra as calculated with the $\{\mathbf{H}_{kl}\}$ are averaged over 200 realizations with the same $k$ resulting for each $k\,{=}\,1,\ldots,n_\mathrm{mat}$ SHM in 10 different spectra with coefficients $\overline{\mathbf{C}}_{kj}$, $j\,{=}\,1,\ldots,10$. 
The network is trained on the $\mathbf{C}_{kj}$ together with the reference $\mathbf{C}_k^{\rm ref}$, i.\,e., only spectra are processed by the network as emphasized by the black frame. The right-hand side (red) sketches how noisy spectra (from either experiment or theory) to be purified are also averaged over 200 realizations before fed into the trained network to retrieve the reference spectrum.} 
\label{fig:sketch}
\end{figure}%
%%%%%%%%%%%%%
To cover the domain of realizable spectra uniformly, we calculate first $4{\times}10^{4}$ reference spectra
\cite{spec}.
Among those we select the $n_{\rm mat}\,{=}\,2{\times}10^{4}$ spectra with the largest mutual difference 
\begin{equation}\label{eq:dist}
D_{ij}=\int\!\d E\,\big|P_{i}(E)-P_{j}(E)\big|.
\end{equation} 
For each member of this subset of reference spectra, we calculate $n_{\rm pul}\,{=}\,500$ fluctuating spectra from noisy pulses generated with the partial-coherence method \cite{pfji+10,suppl} with a different noise realization for each (synthetic) Hamilton matrix. 
Hence, we must propagate about $n_{\rm mat}{\times}n_{\rm pul}\,{=}\,10^{7}$ TDSEs, which takes, however, only a few seconds for a single TDSE thanks to our highly-optimized propagation scheme \cite{suppl}. It includes pre-diagonalization of the Hamilton matrices which saves computing time since one and the same system is propagated for different pulse realizations. 
Finally, we have for each Hamilton matrix \eqref{eq:1d-hama} one reference spectrum  $P_{k}^{\rm ref}(E)$ and $n_{\rm pul}$ noisy spectra $P_{kl}(E)$, i.\,e., a total of $n_{\rm mat}{\times}[n_{\rm pul}{+}1]$ spectra. 

Instead of the individual $P_{kl}(E)$ we use averaged spectra $\overline{P}_{kj}(E)=\frac{1}{m}\sum_{l\in\,\{s_{j}\}}P_{kl}(E)$ for efficient training. To this end we draw a random subset $\{s_{j}\}$ containing $m$ spectra from the $10^{3}$ fluctuating spectra for each SHM and repeat this procedure $\overline n_{\rm pul}$ times to create $j = 1,\ldots,\overline{n}_{\rm pul}$ averaged spectra.
For our application $\overline n_{\rm pul}\,{=}\,10$ and $m\,{=}\,200$ is a good compromise between rugged spectra for smaller $m$ and an increased training effort for larger $m$. 
All spectra are normalized, i.\,e., $\int\!\d E\,P(E)=1$. 

(v) To complete the final step, the parametrization of the spectra for training, we represent
the resulting averaged spectra $\overline{P}_{kj}(E)$ in terms of harmonic oscillator eigenfunctions $\chi_{\kappa}$ as
\begin{equation}\label{eq:spec}
\overline{P}_{kj}(E)=\Big|\sum\nolimits_{\kappa=1}^{n_{\rm bas}}\Cavedef{}_{kj}^{\kappa}\chi_{\kappa{-}1}(E)\Big|^{2},
\end{equation}
with the set $\Cave\,{\equiv}\,\{\Cavedef_{1}\ldots \Cavedef_{n_{\rm bas}}\}$ of coefficients.
A basis size of $n_{\rm bas}\,{=}\,60$ was necessary for the averaged fluctuating spectra,
while using a similar expression for the noise-free spectra $n_{\rm bas}\,{=}\,40$ was sufficient \cite{suppl}. 
The network consists of mapping the coefficients 
$\{\Cave_{kj}\}\to\{\mathbf{C}_{kj}\}$.
%$\{\Cave_{kj},\mathbf{C}_{k}^{\rm ref}\}\to\mathbf{C}_{kj}$.
The training aims at minimizing the difference between the predicted $\mathbf{C}_{kj}$ for the noise-free spectrum and the expected reference spectra $\mathbf{C}_{k}^{\rm ref}$.
%Mathematically, this corresponds to minimizing a cost function, 
%\added{which in our case is given by
%\begin{equation}\label{eq:delta}
%\delta_{\Omega}\equiv\frac{1}{n_{\Omega}}\sum_{j,k=1}^{n_{\Omega}}\big[\mathbf{C}_{kj}-\mathbf{C}_{k}^{\rm ref}
%\big]^{2}.
%\end{equation}
%As discussed below, this function has to be calculated over different data sets $\Omega$, with $n_{\Omega}$ being the size of the specific data set.}

The connection between Hamilton matrices, pulses and electron spectra just outlined is summarized schematically in Fig.\,\ref{fig:sketch}. 

\sec{Building and training the network} 
With $n_{\rm mat}\,{=}\,2{\times}10^{4}$ reference spectra and $\overline{n}_{\rm pul}\,{=}\,10$ averaged noisy ``copies'' of each reference spectrum, we have $n\equiv n_{\rm mat}{\times}\overline{n}_{\rm pul}\,{=}\,2{\times}10^{5}$ pairs available for building the network model. Each pair consists of an averaged noisy spectrum with its respective reference spectrum.
Note that the network operates exclusively on the electron spectra, cf.\,Fig.\,\ref{fig:sketch}. Once trained, it is therefore directly applicable to the experiment which has only access to spectra.

The full data set with $n$ pairs of spectra is split in the ratio 8:1:1 between training ($n_\mathrm{train}\,{=}\,0.8\,n$), validation ($n_\mathrm{val}\,{=}\,0.1\,n$) and test ($n_\mathrm{test}\,{=}\,0.1\,n$) data.
Implemented with the deep-learning library \textsc{Keras} \cite{ch15}, a fully connected feed-forward neural network is used \cite{suppl}. 
The training success and resulting performance of the network as a function of the size of the training data
is quantified with the cost function $\delta$, using the basis representation \eqref{eq:spec} of the spectra, and a more intuitive error $\eps$
\begin{subequations}\label{eq:erros}\begin{align}
\label{eq:delta}
\delta_{\Omega}&\equiv\frac{1}{n_{\Omega}}\sum_{j,k=1}^{n_{\Omega}}\big[\mathbf{C}_{kj}-\mathbf{C}_{k}^{\rm ref}\big]^{2},
\\
\label{eq:eps}
\eps_{\Omega}&\equiv\frac{1}{n_{\Omega}}\sum_{j,k=1}^{n_{\Omega}}D_{kj,k\hspace{0.05em}\rm ref},
\end{align}\end{subequations}
for training ($\Omega\,{=}\,{\rm train}$), validation ($\Omega\,{=}\,{\rm val}$) and test  ($\Omega\,{=}\,{\rm test}$) data set, respectively.
The error $\eps_{\Omega}$ with an upper limit $\eps\,{\le}\,2$ measures the difference $D_{kj,k\rm ref}$, see Eq.\,\eqref{eq:dist}, between a spectrum $P_{kj}(E)$ and its reference spectrum $P_{k}^{\rm ref}(E)$. 
The maximal error $\eps\,{=}\,2$ occurs if the two normalized (i.\,e.\ unit-area) spectra are completely disjunct. 
Both errors \eqref{eq:erros} decay logarithmically as a function of the SHM data size $n$ \cite{suppl}.

%%%%%%%%%%%%%%%%%
\begin{figure}[b]
\centering
\includegraphics[scale=0.65]{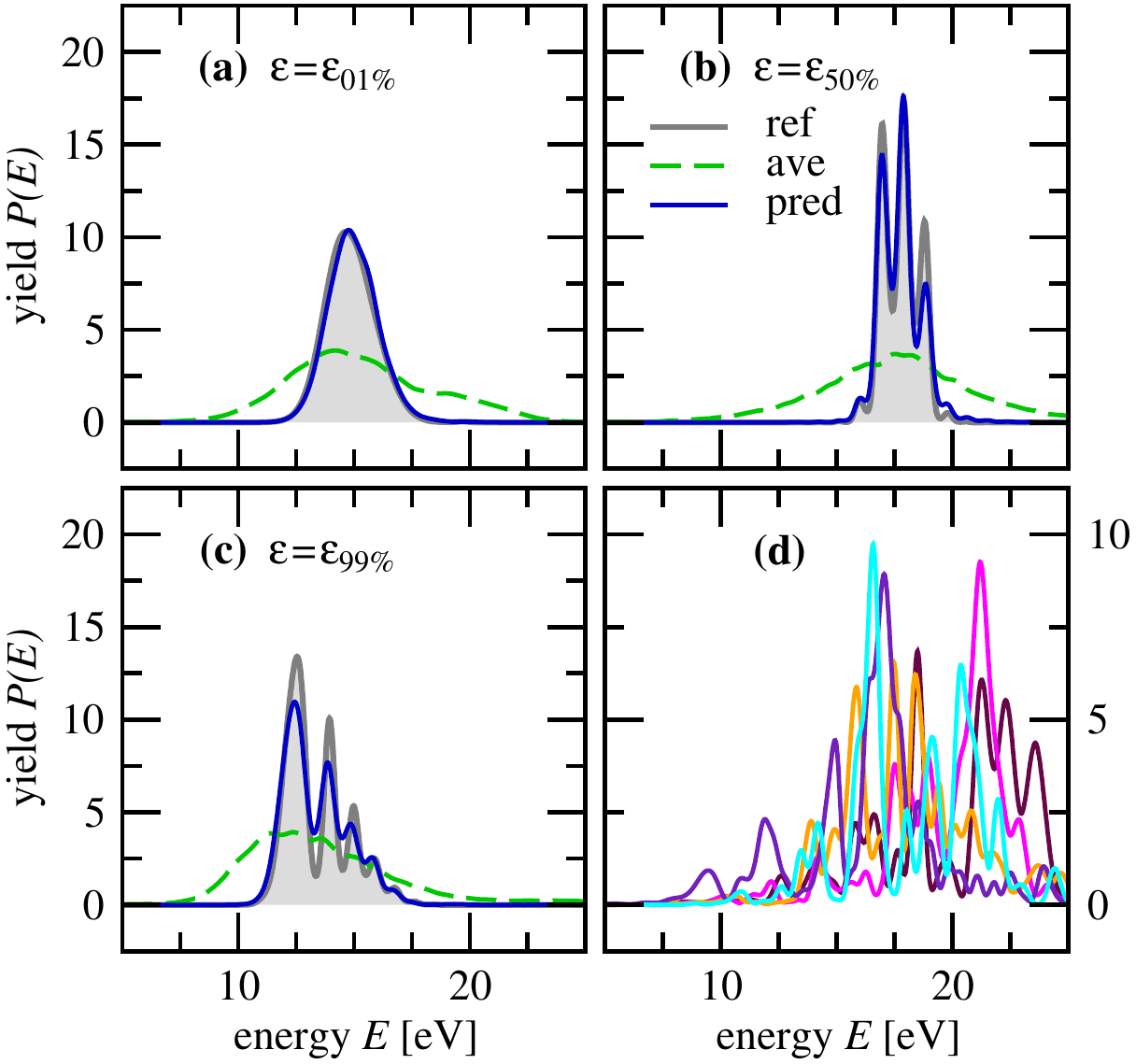}
\caption{Photo-electron spectra from the SHM test-data set.
The average of fluctuating spectra (green-dashed line) and prediction from the network (blue) are compared to the reference (gray and shaded). Panels a--c: Examples with 3 prediction errors $\eps\,{=}\,\eps_{p}$ are shown, with $p$ indicating the percentage of spectra having a smaller error, i.\,e., 99\% of all spectra from the test-data set have a smaller prediction error than the one shown in panel c. 
Panel d: Five single-shot spectra for the Hamilton matrix used in panel b.
}
\label{fig:spec-1d}
\end{figure}%
%%%%%%%%%%%%%

\sec{Purification of spectra from SHMs} We are finally in a position to purify noisy spectra
and do this first with the $n_\mathrm{test}$ SHM-generated spectra the network was not trained on.
Typical snapshots of these spectra are shown in Fig.\,\ref{fig:spec-1d}d.
To get a realistic picture we have selected spectra, cf.\ Fig.\,\ref{fig:spec-1d}\,(a--c), which belong to three groups purified with different residual errors in increasing order: Only 1\% of the spectra have a purification error better than the one shown in Fig.\,\ref{fig:spec-1d}a, the prediction in Fig.\,\ref{fig:spec-1d}b has a median error $\eps=\eps_{50\%}$ such that half of the spectra have a smaller and half of them have a larger prediction error. Finally, only 1\% of the purified spectra have a larger error than the one shown Fig.\,\ref{fig:spec-1d}c. 
The gray-shaded curves provide the reference spectrum $P_k^{\rm ref}(E)$ in each case. 
The simple average (from the test-data set for a specific SHM and field intensity) set is shown as a dashed line.

%%%%%%%%%%%%%%%%%%%%%%%%%%%%
\begin{figure*}[t]\centering
\includegraphics[scale=0.75]{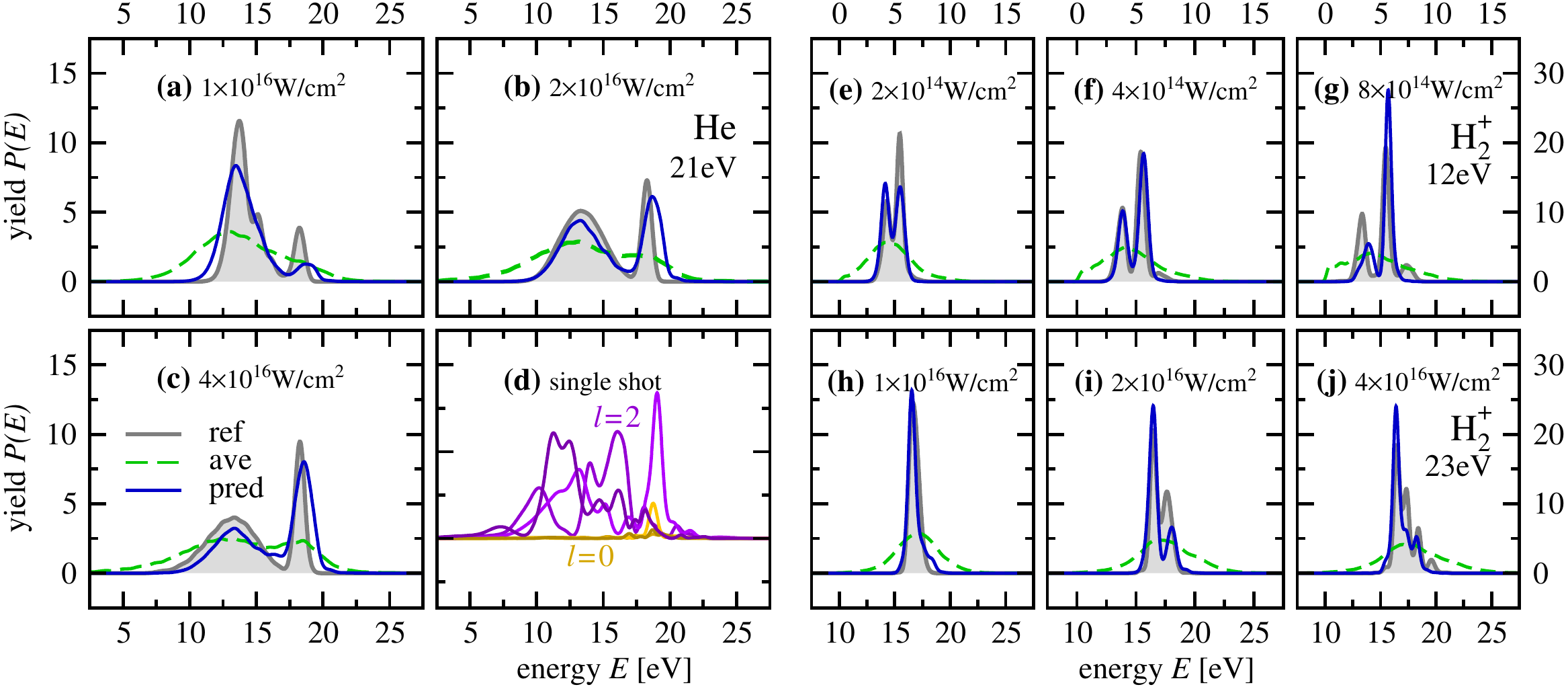}
\caption{Photo-electron spectra for the He atom (a--d) and the H$_{2}{\!}^{+}$ molecule (e--j).
% (d-component, i.\,e.\ $\ell{=}2$) for 3 different intensities, specified in the panels.
For He we show results for one photon-frequency $\omega$ (case {\a} in the text), for H$_{2}{\!}^{+}$ for two different $\omega$ (cases {\b} and {\c} in the text),
whereby each combination is presented for 3 different intensities $I$ (with $\omega$ and $I$ are being specified in the panels).
As in Fig.\,\ref{fig:spec-1d}, averaged and predicted spectra are compared to the reference spectra.
Three single-shot spectra for $I\,{=}\,2{\times}10^{16}$W/cm$^{2}$, as used in panel b, are shown in panel d, separately for the s-channel ($\ell{=}0$, orange-like lines) and the d-channel ($\ell{=}2$, violet-like lines), respectively.
}
\label{fig:spec-3d}
\end{figure*}%
%%%%%%%%%%%%%%

One sees that the purification works quite well, even for a typical ``worst case'' as in Fig.\,\ref{fig:spec-1d}c, where all peaks including the fine structure, appear at the correct energies, despite the fact that none of the features is contained in the averaged spectra.
We also note that spectra of rather different shapes and details of the structure,
from a smooth single peak (Fig.\,\ref{fig:spec-1d}a) over a triple peak (Fig.\,\ref{fig:spec-1d}b) to a fine-structured multi-peak shape (Fig.\,\ref{fig:spec-1d}c), can be purified successfully.
The rather diverse spectra $P_{kl}(E)$ from single fluctuating pulses $f_{l}(t)$, as shown in Fig.\,\ref{fig:spec-1d}d, indicate the strong sensitivity to the pulse profile which is due 
 to Stark shifts and Autler-Townes splittings.
The complete failure of the averaged spectra $\overline{P}_{k}(E)=\sum_{j}\overline{P}_{kj}(E)$ in revealing the reference spectrum $P^\mathrm{ref}_{k}(E)$ is striking. This happens despite the fact that the reference {\it pulse} is retrieved by averaging a sufficient number of fluctuating pulses if created by the partial-coherence method \cite{pfji+10}. The corresponding reference {\it spectrum}, however, is never obtained by averaging the fluctuating spectra since the underlying ionization dynamics is non-linear. The consequence is an intricate mapping between fluctuating spectra and the reference spectrum, which is constructed with the deep neural network.

\sec{Purification of spectra from physical systems} 
So far the successful purification referred to spectra not known to the network, but generated through SHMs which were also used to train the network, only with different parameters. In the following we will apply the network to photo-electron spectra for three cases of two different physical systems:
 (\a) He atoms dominated by 2-photon absorption \cite{sagi+18} and the hydrogen molecule ion H$_{2}^{+}$ ionized by (\b) 2- and (\c) 3-photon processes. These spectra have been obtained in full 3D, for technical details see \cite{suppl}.
In case {\a} (Fig.\,\ref{fig:spec-3d}a--c) the spectra consist of contributions from the s- and d-manifolds, which can be reached by a 2-photon processes, whereby the d-channel clearly dominates (cf.~Fig.\,\ref{fig:spec-3d}d).
For H$_{2}^{+}$, aligned along the laser polarization, either the gerade continuum for case {\b} (Fig.\,\ref{fig:spec-3d}h--j), or the ungerade continuum for case {\c} (Fig.\,\ref{fig:spec-3d}e--g), is considered.
The central frequencies for the laser pulse are chosen according to the transition energies 
$\omega_{\mbox{{\scriptsize\a}}}\,{=}\,E_{\rm 2p}{-}E_{\rm 1s}\,{=}\,20.95\,\mbox{eV}$,
$\omega_{\mbox{{\scriptsize\b}}}\,{=}\,E_{\rm 2\sigma_{\!\rm u}}{-}E_{\rm 1\sigma_{\!\rm g}}{=}\,23.05\,\mbox{eV}$
and
$\omega_{\mbox{{\scriptsize\c}}}\,{=}\,E_{\rm 1\sigma_{\!\rm u}}{-}E_{\rm 1\sigma_{\!\rm g}}{=}\,11.83\,\mbox{eV}$,
respectively.
Fluctuating pulses $f_{l}(t)$ are created as before but we use new random realizations.
As in the training procedure, we have created 10 averaged spectra, which are fed into the trained network. Each one is composed of 200 fluctuating spectra \cite{spec}. The 10 resulting purified spectra from the network are again averaged to arrive at the network's estimate of the reference spectrum. 

We show in Fig.\,\ref{fig:spec-3d} results for three different intensities in the range where few-photon ionization is non-perturbative. As expected from SHM-generated spectra in Fig.\,\ref{fig:spec-1d}, the averaged spectra (green-dashed lines) do not provide sensible information about the reference spectra.
The mapping with the network (blue-solid lines), however, which reveals the respective peak structure of the photo-electron spectra.

Note that the network was neither trained on the 3D helium atom nor on the hydrogen molecule ion, whose spectra
are purified successfully with the network mapping in Fig.\,\ref{fig:spec-3d}. The training of the network was
performed with synthetic data derived from a representative 1D photo-ionization dynamics only, which allowed us to keep the size of the Hamilton matrices small enough to be able to compute the $10^{7}$ TDSEs for a sufficient amount of training data. 
Apparently, although generated from the 1D derived ones, the SHMs represent dynamical systems sufficiently generic such that also realistic 3D spectra from the three rather different processes \a, \b, and \c\ could be purified with
the \emph{same\/} network. Hence, it should also work on experimental spectra, which will be slightly different to the extent to which many-electron effects show up in photo-electron spectra as compared to the present 3D single-active-electron calculations.
To measure reference spectra in a proof-of-principle experiment one could either use seeded FEL pulses \cite{ambe+12,alap+12,alca+13,riab+19} or set up an experiment at a coherent (high-harmonic) source and generate noisy pulses artificially.

To summarize, we have devised a strategy to purify noisy photo-electron spectra, typical for SASE FELs with the help of a deep neural network. While this example was chosen on purpose to be specific, through its design our approach is far more general. Firstly, we have checked \cite{suppl} that other noise models \cite{rosa07,nila12} can be used. 
Secondly, purification could be conditioned on any arbitrary reference pulse. Thirdly, and most importantly,
the systematic introduction of synthetic Hamilton matrices permits to generate a training data set of ample size with reasonable computational effort and renders the trained network applicable for scenarios where it was not trained for. In the present example, we applied the network trained on synthetic dynamics to purify realistic 3D spectra. 
For future work, we would like to point out that noisy pulses driving non-linear processes are actually advantageous, 
since they allow one to obtain the target response over a wide spectral and dynamic range in a single shot, provided one has tools to analyze the resulting spectra.

\sec{Acknowledgements}
This work has been supported by the Deutsche Forschungsgemeinschaft (DFG) through the priority program 1840 ``Quantum Dynamics in Tailored Intense Fields''.

\def\articletitle#1{\emph{#1}.}


\begin{thebibliography}{10}
\bibitem{spec}
We will use the terms ``reference spectrum'' and ``fluctuating spectrum'' to
 refer to a ``spectrum from a reference pulse'' and a ``spectrum from a
 fluctuating pulse'', respectively.
 We never deal with the spectral representation of the pulse itself.
\end{thebibliography}
\end{document}